\documentclass[prl,twocolumn,showpacs,amsmath,amssymb,floatfix,amstex,footinbib]{revtex4-1}

\usepackage{units}
\usepackage{psfrag}

\usepackage[dvips]{graphicx}
\usepackage{bm,pstricks,longtable}

\newcommand{\rmd}{\mathrm{d}}

\newcommand\energy{E}

\newcommand{\new}[1]{#1}

\begin{document}

\title{Scaling of energy spreading
 in strongly nonlinear disordered lattices}

\author{Mario Mulansky} 
\affiliation{\mbox{Department of Physics and Astronomy, Potsdam University, 
  Karl-Liebknecht-Str 24, D-14476, Potsdam-Golm, Germany}}
\author{Karsten Ahnert} 
\affiliation{\mbox{Department of Physics and Astronomy, Potsdam University, 
  Karl-Liebknecht-Str 24, D-14476, Potsdam-Golm, Germany}}
\author{Arkady Pikovsky} 
\affiliation{\mbox{Department of Physics and Astronomy, Potsdam University, 
  Karl-Liebknecht-Str 24, D-14476, Potsdam-Golm, Germany}}

\date{\today}

\begin{abstract}
\noindent 
To characterize a destruction of Anderson localization by nonlinearity, we study the spreading behavior of initially localized states in disordered, strongly nonlinear lattices. Due to chaotic nonlinear interaction of localized linear or nonlinear modes, energy spreads nearly subdiffusively. 
Based on a phenomenological description by virtue of a nonlinear diffusion equation we establish a one-parameter scaling relation between the velocity of spreading and the density, which is confirmed numerically. From this scaling it follows that for very low densities the spreading slows down compared to the pure power law.
\end{abstract}


\maketitle

In linear, disordered, one-dimensional lattices, all eigenmodes are exponentially localized due to Anderson localization \cite{Anderson-58}.
These models have been first suggested for disordered electronic systems \cite{Kramer-MacKinnon-93},
but they are also applicable to a wide range of  wave phenomena (optical, acoustical, etc.) in disordered media~\cite{Sheng-06}. Localization effectively stops spreading of energy in such situations. 

By considering waves of large amplitudes, one faces \textit{nonlinearity} and   meets the question whether  the localization is destroyed \new{due to a nonlinear interaction of eigenmodes}.
Although this question has been addressed numerically \cite{Molina-98,*Kopidakis-Aubry-00,Pikovsky-Shepelyansky-08,Flach-Krimer-Skokos-09,*Veksler-etal-09,Mulansky-Pikovsky-10,*Laptyeva-etal-10}, mathematically~\cite{Bourgain-Wang-07}, and even
experimentally in BECs \cite{Lye-05,*Schulte-05,*Clement-06} and optical waveguides \cite{Schwartz-07,*Lahini-08},
 a full understanding is still elusive. It is easier to understand how nonlinearity destroys localization leading to thermalization and self-transparency in short random lattices~\cite{Mulansky-Ahnert-Pikovsky-Shepelyansky-09,*Tietsche-Pikovsky-08}, than to analyze asymptotic regimes at large times in long lattices. \new{The 
 most striking effect observed in numerical studies is a subdiffusive power-law spreading of energy, lasting until maximally available integration times.  Whether this spreading is an asymptotic state, or transforms into a much weaker logarithmic spreading suggested by some theoretical estimates~\cite{Bourgain-Wang-07,Basko-10}, or even stops, remains a challenging problem.}
 
\new{In this letter we attack the underlying spreading mechanisms by utilizing scaling arguments.}
The concept of scaling has been extremely successful in the understanding of Anderson localization~\cite{Abrahams-etal-79,*Deych-Lisyansky-Altshuler-00,%
*Evers-Mirlin-08}, as well as in descriptions of nonequilibrium phenomena like surface growth~\cite{Barabasi-Stanley-95}.
\new{In this paper we demonstrate that the spreading of energy in strongly nonlinear disordered lattices satisfies scaling relations.}

The starting point is a heuristic description of the spreading by the nonlinear diffusion equation (NDE)~\cite{Mulansky-Pikovsky-10}:
\begin{equation} 
 \frac{\partial \rho}{\partial t} = D\frac\partial{\partial x} \left(\rho^a \frac{\partial \rho}{\partial x}\right),\qquad \text{with} \qquad \int \rho\, \rmd x = E.
 \label{eqn:nl_diff_eq}
\end{equation}
Here $E$ is the total energy (which is conserved) and $\rho$ represents the energy density.  Heuristically, the NDE describes diffusion that appears solely due to nonlinearity, for the nonlinear disordered lattices this can be attributed to a random exchange of energy between modes due to their chaotisation. 
The NDE has a self-similar 
 solution~\cite{Polyanin-Zaitsev-03} describing asymptotic subdiffusive spreading with the edge  of excitation propagating according to:
\enlargethispage{1em}
\begin{equation} \label{eqn:edge_propagation}
  X = \sqrt{2\frac{2+a}a} E^{a/(2+a)}(D(t-t_0))^{1/(2+a)}\;,
\end{equation}
where $t_0$ accounts a time shift depending on the pecularities of the initial state. At the moment, it is not possible to derive the NDE for a particular nonlinear disordered lattice, but one can check if the scaling predicted by Eq.~(\ref{eqn:edge_propagation}) holds. In other words, one can check if a particular nonlinear lattice belongs to a NDE-universality class. The main property of this class is that dependencies on the total energy $E$ and on time  are described by one parameter $a$. This allows us to write the law of edge propagation in a scaled form
\begin{equation} \label{eqn:scaled_spreading}
 \frac{X}{E} \sim \left(\frac{t-t_0}{E^2}\right)^{1/(2+a)}.
\end{equation}
Still, this expression contains an unknown, non-universal constant $t_0$. We get rid of it by considering the local inverse velocity of the edge $dt/dX$ for which holds
\begin{equation}
 \frac{1}{X}\frac{\text{d} t}{\text{d} X}  \sim \left(\frac{E}{X}\right)^{-a}\;. \label{eq:vel}
\end{equation}
We notice that on the r.h.s. the global density of the field $w=E/X$ appears. 
This allows us to rewrite and generalize (\ref{eq:vel}) in the form of an one-parameter scaling relation
\begin{equation}
 a(w)=-\frac{\text{d}\log\frac{1}{X}\frac{\text{d} t}{\text{d} X}}{\text{d} \log w}\;,  \label{eq:vel1}
\end{equation}
where a dependence of the index $a$ on the global density $w$ would indicate deviations from the pure power-law scaling given in eq.~\eqref{eqn:scaled_spreading}. 
\new{Relations (\ref{eqn:edge_propagation},\ref{eq:vel},\ref{eq:vel1}) define the scaling laws to be checked for particular systems; if they are satisfied, then we say that the system belongs to a NDE-universality class.}

We apply the scaling relations (\ref{eqn:edge_propagation},\ref{eq:vel},\ref{eq:vel1}) to 
Hamiltonian lattices with strongly nonlinear coupling:
\begin{equation} \label{eqn:general_hamiltonian}
 H = \sum_k \frac {p_k^2}2 + W\omega_k^2 \frac {q_k^\kappa}\kappa + \frac \beta \lambda (q_{k+1} - q_k)^\lambda,
\end{equation} 
with $\kappa \geq 2$, $\lambda > 2$ being the powers of the local and the coupling potential, respectively. Random frequencies
$\omega_k$  of the $k$-th oscillator are chosen to be independent  random numbers uniformly distributed on the interval $(0,1)$. Such lattices without local potential (${W=0}$) possess traveling compact waves (compactons)~\cite{Ahnert-Pikovsky-09}. With a regular local potential the compactons are long-living objects dominating the energy spreading and making it nearly ballistic, while the presence of disorder surpresses them and makes spreading subdiffusive as will be presented below.
Parameter~$W$ describes the disorder strength whereas $\beta$ governs the nonlinear coupling.
However, in the case $\kappa \neq \lambda$ by applying a transformation
$ q_k  \to W^\alpha \beta^{-\alpha} q_k$, 
$ p_k \to W^{\lambda\alpha/2}\beta^{-\kappa\alpha/2} p_k$ and
$ H   \to W^{\lambda\alpha}\beta^{-\kappa\alpha} H$, 
with $\quad \alpha = 1/(\lambda-\kappa)$,
 we can set parameters $W$ and $\beta$ to one. Thus, the  only relevant parameter is the total energy~$E$.
Hence, for $\kappa \neq \lambda$, varying the disorder strength~$W$ or the nonlinear strength~$\beta$ is strictly equivalent to appropriate changes of the energy~$E$ in this system.

For the special case $\kappa = \lambda$ the system exhibits full energy scaling which is seen from the invariance of the equations under the following transformations:
\begin{equation} \label{eqn:energy_scaling}
\begin{aligned}
 q &\to q' = \gamma q,&\qquad
 p & \to p' = \gamma^{\kappa/2} p, \\
 t &\to t' = t/\gamma^{\kappa/2-1}, &\qquad
 E & \to E' = \gamma^{\kappa} E.
\end{aligned}
\end{equation}
Furthermore, we can replace ${q_k \to W^{-1/\kappa}q_k}$ and ${\beta \to \beta/W}$ and finally end up at the Hamiltonian (\ref{eqn:general_hamiltonian}) with $W=1$ and $\lambda=\kappa$.
Now the only parameter is $\beta$, which describes the relative strength of the local and the coupling potentials.

The case ${\kappa=\lambda}$ is special because here we can establish an exact relation between the  parameter of NDE $a$ and the nonlinearity index $\kappa$.
Indeed, excluding $\gamma$ from the scalings of time and energy in (\ref{eqn:energy_scaling}) we obtain $t\sim E^{\frac{2-\kappa}{2\kappa}}$. Comparing this with the scaling   $t-t_0 \sim X^{2+a} \energy^{-a}$ that follows from Eq.~(\ref{eqn:edge_propagation}), we find $a=\frac{\kappa-2}{2\kappa}$. 
\new{(Note, that we have not derived the NDE from the Hamiltonian, but solely 
use scaling arguments to find the exact correspondance between $a$ and $\kappa$. Validity of the approach has still to be checked numerically below.)}
From the expression for $a$, we find the spreading law:
\begin{equation}
X\sim (t-t_0)^{\frac{2\kappa}{5\kappa-2}}\;.
\label{eq:sp1}
\end{equation}
For this lattice with homogeneous nonlinearity (in the sense that local and coupling potenials have the same nonlinearity index~$\kappa$) the one parameter scaling predictions (\ref{eqn:scaled_spreading},\ref{eq:vel},\ref{eq:vel1}) are trivially fulfilled with index $a$ being independent on density $w$, as the energy dependence follows from exact rescaling. For lattices with nonhomogeneous nonlinearity $\lambda\neq \kappa$ the density dependence (\ref{eqn:scaled_spreading},\ref{eq:vel},\ref{eq:vel1}) is nontrivial.

\begin{figure}[t]
  \centering
  \includegraphics[height=0.43\textwidth, angle=270]{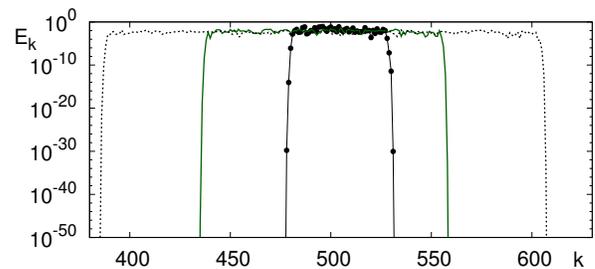}
  \caption{Energy profiles $E_k$ for ${\kappa=2}$ and $\lambda=4$ of initially localized states at different times $t = 10^4,\, 10^6,\, 10^8$ (inner to outer, one specific disorder realization, $E=1$). }
  \label{fig:wf_times}
\vglue-1em
\end{figure}

For the Hamiltonian \eqref{eqn:general_hamiltonian}, 
 defining the local energy at site $k$ as
\begin{equation*}
\begin{aligned}
 E_k &= \frac {p_k^2}2 + W\omega_k^2 \frac {q_k^\kappa}\kappa + \frac \beta {2\lambda} \left((q_{k+1} - q_k)^\lambda + (q_k - q_{k-1})^\lambda \right),
\end{aligned}
\end{equation*}
we can interprete it as the distribution $\rho(k,t) = E_k$, time evolution of which we compare with predictions of NDE. The numerically obtained profiles of $E_k(t)$ are depicted in Fig.~\ref{fig:wf_times}.  The peculiar property of strongly nonlinear lattices of type (\ref{eqn:general_hamiltonian}) is that the field has very sharp edges: one can estimate that the tail decays superexponentially (like for compactons in systems without disorder~\cite{Ahnert-Pikovsky-09}). This corresponds well to the property of the self-similar solution of the NDE to have a sharp edge, and allows us to compare predictions of the scaling theory with the numerics for the lattices.

\begin{figure}[t]
   \centering
   \includegraphics[height=0.43\textwidth,angle=270]{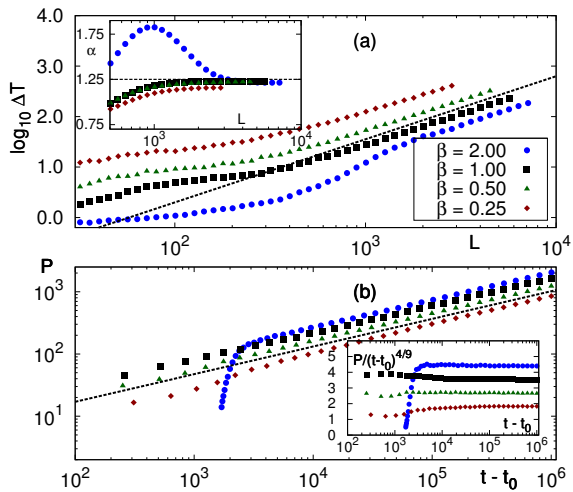}
\vglue -0.5em
\caption{(color online) Excitation times~$\Delta T(L)$ (a) and participation number~$P(t-t_0)$ (b) of spreading states in the hamiltonian lattice \eqref{eqn:general_hamiltonian} with ${\kappa=\lambda=4}$.
Values $t_0$ are adjusted to have a maximally extendended range of power law behavior.
The dotted lines correspond to the scaling results ${a=(\kappa-2)/(2\kappa)=1/4}$ and have slopes $5/4$ for $\Delta T$ and $4/9$ for $P$.
The inset in (a) shows the instantanious time dependent exponent $\alpha$ from $\Delta T \sim L^\alpha$ saturating at the expected value $\alpha = 5/4$.
The inset in (b) shows the rescaled spreading due to the scaling expectation ${P/(t-t_0)^{4/9}}$ vs.\ $t-t_0$.
}
\label{fig:dn_P_nl_do}
\end{figure}

We start with testing the approach by applying extensive numerical simulations to the case ${\kappa=\lambda=4}$ where theory (\ref{eq:sp1}) predicts  ${a=1/4}$. 
We integrated the equations of motion by means of a fourth order symplectic Runge-Kutta method~\cite{McLachlan-95}. In all simulations in this paper we used time steps in the range 0.01 --- 0.1 that assured conservation of energy with accuracy $\sim 10^{-9}$, presented results are averages over hundreds of realizations of disorder.
Our main quantity of interest is the time $\Delta T(L)$ required to excite the next oscillator with $L$ already beeing excited. This quantity can then be interpreted as the inverse velocity of the edge that enters Eq.~(\ref{eq:vel},\ref{eq:vel1}), where the transition $L\to L+1$ implies $\text{d}X=1$.
We defined an oscillator to be excited if $E_k > 10^{-50}$.
(Due to the sharp edges of the states this is a reasonable, though arbitrary, value; changing this threshold to, say, $10^{-20}$ or $10^{-100}$ produces similar results.)
To obtain the mean value of $\Delta T(L)$, we averaged, for each $\beta$, $\log(\Delta T)$ for each $L$ over disorder realizations.
For a better visual display of the excitation times on a logarithmic scale we also averaged $\Delta T$ over neighbouring sites $L$. The results are shown in Fig.~\ref{fig:dn_P_nl_do} (a), at large $L$ they are  in perfect correspondance with the theoretical prediction ${\Delta T \sim L^{5/4}}$.

We compare these results with a more traditional approach, where the width of the field distribution is averaged over realizations of disorder for fixed moments of time. We have calculated two measures of the width -- the squared mean displacement $(\Delta k)^2$ and the participation number of the energy distribution $P$, as already used in literature~\cite{Kopidakis-08,Pikovsky-Shepelyansky-08,Skokos-Flach-10,*Laptyeva-etal-10}.
However, both quantities behave identically in our numerical simulations, hence we show only the participation number exemplarily representing the spatial extent, defined as
$  P^{-1} =  \sum_k (E_k/E)^2$.
Obviously, we have ${P \sim \Delta k \sim L\sim X}$.
Values of~$P$ have been averaged over increasing time windows and disorder realizations, the results are shown in
Fig.~\ref{fig:dn_P_nl_do} (b). 
These time evolutions fit nicely Eq.~(\ref{eqn:scaled_spreading}) with the theoretically predicted value $a=1/4$.
\new{Hence, we have found the NDE to reproduce the correct spreading behavior of this nonlinear disordered lattice. Based on the assumption of the validity of the NDE we were able to calculate the correct spreading exponent analytically. Note, that no further assumptions or parameter fits were required in this case.}


Although the two methods used -- propagation times $\Delta T(L)$ and mean packet widths $P(t)$ (or, equivalently, $L(t)$) -- appear to be nearly equivalent, the former one has two clear advantages. First, it does not possess an arbitrary parameter $t_0$ as the time differences are calculated. The second, and more important, advantage is that by fixing the system length $L$ in the averaging over disorder we in fact fix the characteristic energy density~$w$. On the contrary, by averaging the width at a certain time we do not fix the energy density, as at a given time the variations of the density in different realizations of disorder may be enormous. For the model with homeogeneous nonlinearity $\kappa=\lambda$ this is not essential as the time scales with energy in a trivial manner. For the nonhomogeneous Hamiltonians, to be considered below, this is crucial, as one expects the properties of the spreading to depend intrinsically on the energy density, but not explicitely on time.

\begin{figure}[t]
    \includegraphics[width=0.45\textwidth]{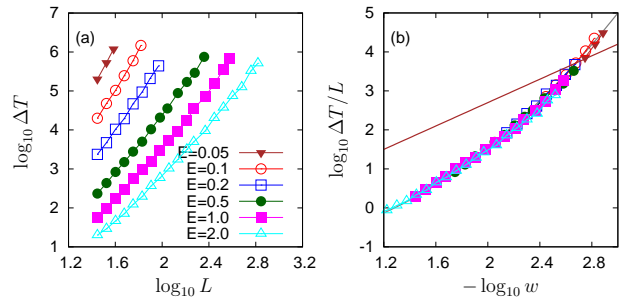}
   \caption{(color online) Panel (a): Excitation times $\Delta T$ of spreading states for several energies $E=0.05\dots 2$. The results shown in this plot are averaged over logarithm intervals on $L$ (see text).
Panel (b) shows the rescaled quantities  $\log_{10}(\Delta T/L)$ vs.~$-\log_{10}w$ with $w=E/L$ according to Eq.~\eqref{eq:vel}; here the solid line is the slope $a(w)$ of the fitting parabolic curve according to Eq.~\eqref{eq:vel1}.
  \label{fig:exc_times}
  }
\vglue-1em
\end{figure}

Above, we have checked the approach on the homogeneous Hamiltonian model with $\kappa=\lambda$, where the scaling with the energy is trivial. Now we apply our method to the mostly nontrivial nonhomogeneous case $\kappa\neq\lambda$, where the theory based on the NDE predicts one-parameter scaling laws (\ref{eqn:scaled_spreading},\ref{eq:vel},\ref{eq:vel1}). More precisely, we focus on the case ${\kappa=2}$ and ${\lambda=4}$ which resembles the widely studied problem of the discrete Anderson nonlinear Schr\"odinger equation (DANSE): its Hamiltonian in the eigenmode representation also possesses a 
quadratic local disorder term and a nonlinear fourth order mode-to-mode coupling.

\begin{figure}[t]
  \includegraphics[height=0.45\textwidth, angle=270]{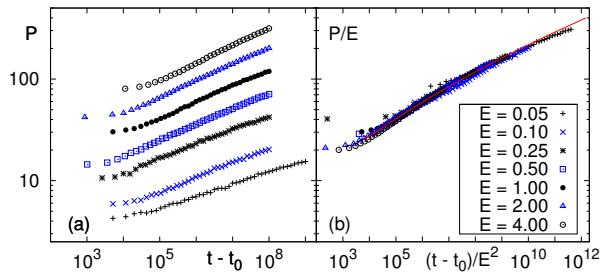} 
  \caption{(color online) (a): Participation Number~$P$ of initially localized states for several different energies $E=0.05\dots 4$ (cf.~fig.~\ref{fig:exc_times}). 
Panel (b) shows the rescaled quantities $P/E$ vs.~${(t-t_0)/E^2}$ according to eq.~\eqref{eqn:scaled_spreading}. The red (grey) line in (b) visualizes the analytic expectation ${P/E \sim X/E = 1/w}$ obtained from Eq.~\eqref{eq:vel} and \eqref{eq:appra}.
}
  \label{fig:p_time}
\vglue-1em
\end{figure}

First, we investigated the excitation times~$\Delta T(L)$ -- the results are shown in Fig.~\ref{fig:exc_times} (a). After performing the scaling according to \eqref{eq:vel}, all the curves collapse to one as seen in panel (b) of Fig.~\ref{fig:exc_times}. The same approach applied to the participation number also leads to a collapse of data when the scaling representation \eqref{eqn:scaled_spreading} is used (see Fig.~\ref{fig:p_time}). The collapse of data for different energies proves numerically that the one parameter scaling suggested by Eqs.~(\ref{eq:vel},\ref{eq:vel1}) works nicely for the strongly nonlinear lattice~\eqref{eqn:general_hamiltonian}.
\new{The validity of this scaling means that asymptotically the spreading in such systems is governed solely by the average energy density~$w$. This was assumed in most of the previous works on this topic \cite{Pikovsky-Shepelyansky-08,Skokos-Flach-10,*Laptyeva-etal-10}, but we present here the first direct numerical evidence of this.}

Fig.~\ref{fig:exc_times} (b) shows that within the studied range of two decades of variations of the density, parameter $a(w)$ (the slope of the curve in Fig.~\ref{fig:exc_times}) is not a constant, but a growing function of inverse density. In particular, data in Fig.~\ref{fig:exc_times} can be well fitted with a linear dependence
\begin{equation}
 a(w)\approx -0.3 - 1.5\log_{10} w
\label{eq:appra}
\end{equation} 
  what corresponds to a parabolic fit for the dependence of $\log_{10}\Delta T/L$ on $\log_{10}w$.
This means that the spreading of energy in the lattice is not a pure power law, but slows down as the density decreases. In the intermediate range of densities the parameter is close to $a=3$, what means that in this range the width of the wave packet spreads as $X\sim t^{1/5}$, i.e. with approximately the same index as found   numerically for the nonlinear Schr\"odinger lattice with disorder in Refs.~\cite{Pikovsky-Shepelyansky-08,Flach-Krimer-Skokos-09,Laptyeva-etal-10}. However, for this model no numerical slowing down of the spreading have been reported, although a recent theoretical estimation in~\cite{Basko-10} gives a sub-power law  asymptotics   $\log P \sim  \log^{1/3} t $. We stress that an application of 
the one parameter scaling  (\ref{eq:vel1}) with the empirical law (\ref{eq:appra}) also demonstrates a good agreement for the participation number as seen in Fig.~\ref{fig:p_time}  (compare the line in panel (b) with the markers).

In conclusion, we have studied subdiffusive spreading of energy in Hamiltonian lattices with both linear and nonlinear disorder and nonlinear nearest neighbour coupling. Our main result is the one-parameter scaling relation, Eq.~\eqref{eq:vel1}, which relates the average velocity of spreading with the energy density. This scaling relation is motivated by using the nonlinear diffusion equation as a phenomenological model for the macroscopic properties of spreading. We studied in details two sets of exponents on the Hamiltonian lattice, $\kappa=\lambda=4$ and $\kappa=2\,,\;\lambda=4$. In the first case there is no nontrivial dependence on the energy, what allowed us to find the spreading index analytically and to confirm it numerically. 
This agreement of analytical and numerical results is not surprising, but still remarkable because it shows that indeed the NDE is an appropriate framework to approach systems with disorder and nonlinearity.
The latter case of linear disorder and nonlinear coupling is mostly nontrivial, here our approach gave a density dependent index $a(w)$, that in a large range of densities is close to $a\approx 3$ but grows as density decreases in course of spreading.
This dependence $a(w)$ has not been observed before (e.g.~in the DANSE model), and it is a first indication of a deviation of the spreading from the perfect subdiffusive power law.

While we studied in details the strongly nonlinear Hamiltonian lattices, it remains a challenge to extend the results to lattices with linear coupling terms, e.g. on the nonlinear disordered Schr\"odinger lattice. The main issue here is that for latter situations one cannot define a sharp edge of the spreading wave packet, thus the calculation of the edge velocity, entering the scaling relation~(\ref{eq:vel1}) is problematic. Further studies on lattices of nonlinear oscillators coupled by a nonlinearity of different order, e.g.~$\kappa=4$, $\lambda=6$, are currently pursued.

We thank S.~Flach, D.~Shepelyansky, S.~Fishman and S.~Aubry for useful discussions. Support from DFG project PI220/12 is acknowledged.

\vglue-1em


%

\end{document}